\documentclass[a4paper,10pt,conference]{IEEEtran}

\usepackage[dvipdf]{graphicx}
\usepackage{epsfig}
\usepackage{color}

\usepackage{amsmath}
\usepackage{amssymb}

\begin{document}

\title{\huge{Differential Diversity Reception of MDPSK over Independent Rayleigh Channels with Nonidentical Branch 
Statistics and Asymmetric Fading Spectrum}}

\author{
\authorblockN{Hua Fu and Pooi Yuen Kam}
\authorblockA{ECE Department, National University of Singapore \\ 
Singapore 117576, Email: \{elefh, elekampy\}@nus.edu.sg}
\vspace{-20pt}
}

\maketitle

\begin{abstract}
This paper is concerned with optimum diversity receiver structure and its performance analysis of differential phase shift keying (DPSK) with differential detection over nonselective, independent, nonidentically distributed, Rayleigh fading channels. 
The fading process in each branch is assumed to have an arbitrary Doppler spectrum with arbitrary Doppler bandwidth, but to have distinct, asymmetric fading power spectral density characteristic. Using 8-DPSK as an example, the average bit error probability (BEP) of the optimum diversity receiver is obtained by calculating the BEP for each of the three individual bits. The BEP results derived are given in exact, explicit, closed-form expressions which show clearly the behavior of the performance as a function of various system parameters.
\end{abstract}


\section{INTRODUCTION}
The receiver structure and bit error probability (BEP) performance of differential phase shift keying (DPSK) with differential detection over nonselective, independent and identically distributed (i.i.d.), Rayleigh fading channels with combining diversity reception have been well known in the literature [1]$-$[4]. 
However, reaserch shows that in some practical systems, the independent, non-identically distributed (i.n.i.d.) channel model is more accurate [5], [6]. In i.n.i.d. channel, the fading processes and possibly the additive, white Gaussian noise (AWGN) on the diversity branches have non-uniform power profiles which are distinct from one another. The effect of the nonidentical diversity branch statistics on the receiver structure is studied in [7]. Recently, based on the maximum {\em a posteriori} probability (MAP) criterion, an explicit structure of the optimum combining differential receiver and  a complete set of closed-form BEP expressions and their Chernoff upper bounds, for 2-, 4- and 8-DPSK, both with optimum combining reception and suboptimum combining reception, are derived in [8]$-$[10]. The purpose of this paper is to provide a further extension. The results derived in this paper, together with those in [8]$-$[10], form a benchmark counterpart to the classic ones for the i.i.d. channel given in [1]$-$[4].   

In a Rayleigh channel, the fading gain is usually modeled as a zero-mean, stationary, complex, Gaussian random process. The most widely accepted model [1]$-$[10] is that the spectrum of the fading process over each diversity branch is symmetric around the carrier so that the quadrature processes are independent of each other. This assumption is valid for various fading spectra. For example, see [11] and its references. However, in some fading environments such as the land mobile channel with Jakes model [12], the Doppler spectrum becomes asymmetric when the multipath signals are absorbed by obstacles or the propagation environment is characterised by
directional non-isotropic scattering [13]$-$[15]. Thus, it is of great practical importance to take account of the effect of the asymmetric fading spectrum on the receiver structure and the performance analysis of differentially detected DPSK over i.n.i.d. channels, the topic of this paper.     

The paper is orgainzed as follows. In Section II, the signal model is introduced and different optimum diversity receivers are derived for different Rayleigh fading scenarios (see eqs. (17)$-$(20) below). In Section III, we use 8-DPSK as an example to study the BEP performance. Here, the average BEP of the optimum diversity receiver is obtained by calculating the BEP for each of the three individual bits. The results are given in exact, explicit, closed-form expressions which show clearly the behavior of the performance as a function of signal-to-noise ratio (SNR), fading correlation coefficient, and diversity order. Section IV presents numerical examples. Throughout this paper, overhead $\sim$ denotes a complex quantity, superscript $*$ will denote its conjugate, ${\rm{E}}$ is the ensemble average operator, $\delta$ represents the Kronecker delta, and $[\cdot]^T$ denotes transposition of the vector and matrix.

\section{SIGNAL MODEL AND RECEIVER STRUCTURE}

With space diversity reception over $L$ frequency nonselective, i.n.i.d., Rayleigh fading 
branches with AWGN, the received signal over the $i$th branch, $i=1, 2, \cdots, L,$ during the $k$th symbol interval $kT \leq t< (k+1)T$ is given, after matched filtering and sampling at time $t=(k+1)T$, 
by the statistic $\tilde{r}_{i}(k)$, where
\begin{equation}
\tilde{r}_{i}(k)=E_{s}^{1/2}e^{j\phi(k)}\tilde{c}_{i}(k)+\tilde{n}_{i}(k).
\end{equation}
Here, $E_{s}$ is the energy per symbol, and for DPSK, $\phi(k)$ is the data-modulated phase with Gray encoding of bits onto the phase transition $\Delta\phi(k)=\phi(k)-\phi(k-1)$. The $k$th data symbol is conveyed in $\Delta\phi(k)$. We assume here that all symbol points are equally likely.
In (1), a rectangular data pulse shape $g(t)$, where $g(t)=1/\sqrt{T}$ for $0 \le t <T$ and zero elsewhere, is assumed so that each matched filter has a rectangular low-pass-equivalent impulse response $h_{i}(t)=g(T-t)$ for all $i$. Thus,
the filtered noise $\tilde{n}_{i}(k)$ is given by
\begin{equation}
\tilde{n}_{i}(k)=\int_{kT}^{(k+1)T}\frac{\tilde{n}_{i}(t)}{\sqrt{T}}\,dt.
\end{equation}
Here, $\left\{\tilde{n}_{i}(t)\right\}_{i=1}^{L}$ is a set of i.n.i.d., lowpass, 
complex AWGN processes with ${\rm{E}}\left[\tilde{n}_{i}(t)\right]=0$ and
$
{\rm{E}}[\tilde{n}_{i}(t)\tilde{n}_{i}^{*}(t-\tau)]=N_i \delta(\tau)
$
so that $\left\{\tilde{n}_{i}(k)\right\}_{k}$ is a sequence of zero-mean, complex Gaussian variables with covariance function for each branch $i$
\begin{equation}
{\rm{E}}[\tilde{n}_{i}(k)\tilde{n}_{i}^{*}(j)]=N_i\, \delta_{kj}
\end{equation}
The multiplicative distortion $\tilde{c}_{i}(k)$ in (1) is given by
\begin{equation}
\tilde{c}_{i}(k)=\int_{kT}^{(k+1)T}\frac{\tilde{c}_{i}(t)}{T}\,d t.
\end{equation}
Here, $\big\{\tilde{c}_{i}(t)=a_i(t)+jb_i(t)\big\}_{i=1}^{L}$ is a set of i.n.i.d., lowpass, zero-mean, stationary,
complex, Gaussian random processes. Each $\tilde{c}_{i}(t)$ represents the complex gain due to frequency nonselective Rayleigh fading of the $i$th branch. For asymmetric spectrum in each $i$, the inphase fading process $a_i(\cdot)$ and the quadrature phase fading process $b_i(\cdot)$ are generally correlated. At any time instant $t$, however, $a_i(t)$ and $b_i(t)$ are always uncorrelated. With reference to Fig. 1, it is shown in [16] that the covariance function of $a_i(\cdot)$ and $b_i(\cdot)$ can be obtained as
\begin{subequations}
\begin{align}
{\rm E}[a_i(t)b_i(t)]&=0\\
{\rm E}[a_i(t-\tau)a_i(t)]&={\rm E}[b_i(t-\tau)b_i(t)]=R_i(\tau)\\
{\rm E}[a_i(t)b_i(t-\tau)]&=-{\rm E}[b_i(t)a_i(t-\tau)]=Q_i(\tau)
\end{align}
\end{subequations}
Note that if the spectrum of each $\tilde{c}_i(t)$ is symmetric,  the processes $a_i(\cdot)$ and $b_i(\cdot)$ will be independent (i.e., we have $Q_i(\tau)=0$) with the same covariance function $R_i(\tau)$. 

Letting $\tilde{c}_{i}(k)=a_i(k)+jb_i(k)$, it follows from (4) and (5) that both $\{a_i(k)\}_k$ and $\{b_i(k)\}_k$ are sequences of zero-mean, real-valued, Gaussian random variables with 
\begin{subequations}
\begin{align}
{\rm E}[a_i(k)b_i(k)]&=0\\
{\rm E}[a_i(k-l)a_i(k)]&={\rm E}[b_i(k-l)b_i(k)]=C_i(l)\\
&=\int_{kT}^{(k+1)T}\int_{(k-l)T}^{(k+1-l)T}\frac{R_i(u-v)}{T^2}du\, dv
\nonumber\\
{\rm E}[a_i(k)b_i(k-l)]&=-{\rm E}[b_i(k)a_i(k-l)]=D_i(l)\\
&=\int_{kT}^{(k+1)T}\int_{(k-l)T}^{(k+1-l)T}\frac{Q_i(u-v)}{T^2}du\, dv 
\nonumber
\end{align}
\end{subequations}
Thus, the covariance matrix can be obtained as
{\setlength\arraycolsep{2pt}
\begin{eqnarray}
\Gamma_i&=&{\rm E}\left[
\left[ \begin{array}{c}
a_i(k)\\
a_i(k-l)\\
b_i(k)\\
b_i(k-l)
\end{array} \right]
\;
\Big[a_i(k)\;a_i(k-l)\;b_i(k)\;b_i(k-l)\Big]\right]
\nonumber\\
&=&
\left[\begin{array}{cccc}
C_i(0) & C_i(l) & 0 & D_i(l)  \\
C_i(l) & C_i(0) & -D_i(l) & 0 \\
 0   & -D_i(l)& C_i(0) & C_i(l)\\
D_i(l) & 0 & C_i(l) & C_i(0)
\end{array} \right]
\end{eqnarray}}
For each $i \,$, $\tilde{c}_{i}(k)$ and $\tilde{n}_{i}(k)$ are mutually independent. 
For $i \ne j,$ $\; \{\tilde{c}_{i}(k),\tilde{n}_{i}(k)\}$ are independent of $\{\tilde{c}_{j}(k), \tilde{n}_{j}(k)\}$. 
The diversity branches are nonidentical since the covariance functions $R_i(\tau)$, $Q_i(\tau)$ and $N_i\delta(\tau)$ depend on the branch index $i$.
For convenience of later application, the following parameters are defined. The fading correlation coefficient at the matched filter output over a symbol interval of $T$ for the $i$th diversity branch is defined as 
\begin{eqnarray}
\tilde{\rho}_i=\frac{{\rm{E}}[\tilde{c}_{i}(k)\tilde{c}_{i}^{*}(k-1)]}
{\sqrt{{\rm{E}}\big[\vert\tilde{c}_{i}(k)\vert^2\big]}
{\sqrt{{\rm{E}}\big[\vert\tilde{c}_{i}(k-1)\vert^2\big]}}}=
\frac{C_{i}(1)-jD_i(1)}{C_{i}(0)}
\end{eqnarray}
From (8), we note that $\tilde{\rho}_i$ is a complex quantity. It is a measure of the fluctuation rate of the channel fading process. The mean received SNR per symbol over the $i$th branch is defined as 
\begin{equation}
\gamma_{i}=\frac{{\rm E}\big[\vert E_{s}^{1/2}e^{j\phi(k)}\tilde{c}_{i}(k)\vert^2\big]}{N_i}
=\frac{2E_{s}C_i(0)}{N_i}
\end{equation}
We consider 2-, 4- and 8-DPSK with Gray encoding of bits onto $\Delta\phi(k)$ as shown in [4, Fig.1] for 4- and 8-DPSK, the mean received SNR per bit $\gamma^b_i$ is given by $\gamma^b_i=\gamma_i$ for 2-DPSK, $\gamma^b_i=\gamma_i/2$ for 4-DPSK, and $\gamma^b_i=\gamma_i/3$ for 8-DPSK.

Using the MAP criterion, the aim of the receiver is to determine from the received signals $\{\tilde{r}_{i}(k),\,\tilde{r}_{i}(k-1)\}_{i=1}^{L}$ 
which one of the possible values ${2\pi m}/{M}$, $m=0, 1, \cdots, M-1$, of the phase difference $\Delta \phi (k)$ has maximum probability of occurrence. Following [9], it can be shown that MAP detection is equivalent to maximum log-likelihood detection. Specifically, based on $\{\tilde{r}_{i}(k),\,\tilde{r}_{i}(k-1)\}_{i=1}^{L}$, we decide that $\Delta \phi (k)= {2\pi n}/{M}$ whenever the log-likelihood function
\begin{eqnarray}
{\rm{log}}\Psi_m = \sum_{i=1}^{L} {\rm{log}}\left\{ p\left[\tilde{r}_{i}(k) {\Big{\vert}} \tilde{r}_{i}(k-1), \Delta \phi (k)= \frac{2\pi m}{M} \right]\right\}
\end{eqnarray} 
is maximized for $m=n$. 

To proceed with evaluating (10), we need to verify that $\tilde{c}_{i}(k)=a_i(k)+jb_i(k)$ and $\tilde{c}_{i}(k-1)=a_i(k-1)+jb_i(k-1)$ are {\em jointly complex Gaussian}. By being {\em jointly complex Gaussian}, it means that if ${\bf \widetilde{x}}={\bf x_R}+ j{\bf x_I}$ and ${\bf \widetilde{y}}={\bf y_R}+ j{\bf y_I}$ are two column complex random vector, then $[{\bf x_R}^T\;{\bf y_R}^T\;{\bf x_I}^T\;{\bf y_I}^T]^T$ has a real multivatiate Gaussian probability density function (PDF), and furthermore, if ${\bf u}=[{\bf x_R}^T\;{\bf y_R}^T]^T$ and ${\bf v}=[{\bf x_I}^T\;{\bf y_I}^T]^T$, then the real covariance matrix of $[{\bf u}^T\;{\bf v}^T]^T$ has a special form given in [18, Theorem 15.1] that satisfies Goodman's theorem [19]. After careful examination, it follows from (7) that $\tilde{c}_{i}(k)$ and $\tilde{c}_{i}(k-1)$ are indeed {\em jointly complex Gaussian}\footnote{We also call them the {\it proper} complex Gaussian random variables [17].}. Thus, conditioned on $\tilde{c}_i(k-1)$, $\tilde{c}_i(k)$ is conditionally complex Gaussian with mean [18]
\begin{equation}
{\rm{E}}\left[\tilde{c}_i(k)\vert \tilde{c}_i(k-1)\right]=\tilde{\rho}_i\,\tilde{c}_i(k-1)
\end{equation}
and variance
\begin{eqnarray}
&&{\rm{E}}\left\{\big\vert \tilde{c}_i(k)-{\rm{E}}[\tilde{c}_i(k)\vert\tilde{c}_i(k-1)]\big\vert^2 \Big{\vert}\tilde{c}_i(k-1) \right\}
\nonumber\\
&&\quad =2C_i(0)-2\frac{C_i^2(1)+D_i^2(1)}{C_i(0)}
\end{eqnarray}
Moreover, conditioned on the vector $[a_i(k-1) \; b_i(k-1)]^T$, the vector $[a_i(k) \; b_i(k)]^T$ is conditionally Gaussian with covariance matrix given by
\begin{equation}
\Omega_i=
\left[ \begin{array}{cc}
C_i(0)-\frac{C_i^2(1)+D_i^2(1)}{C_i(0)} & 0  \\
0 & C_i(0)-\frac{C_i^2(1)+D_i^2(1)}{C_i(0)} 
\end{array} \right]
\end{equation} 
which is a diagonal matrix. This shows that ${\rm{Re}}[\tilde{c}_i(k)\vert\tilde{c}_i(k-1)]$ and ${\rm{Im}}[\tilde{c}_i(k)\vert\tilde{c}_i(k-1)]$ are independent.

Applying (11), (12) and (13) to (10), we obtain
\begin{eqnarray}
&&\frac{1}{2}{\rm{log}}\Psi_m =\zeta + 
\\
&&{\rm Re}\left[\sum_{i=1}^{L}\frac{2E_s\,[C_i(1)+jD_i(1)]\; e^{-j\frac{2\pi m}{M}}\;\tilde{r}_{i}(k) \tilde{r}_{i}^{*}(k-1)}{[2E_{s}C_i(0)+N_i]^2-4E_s^{2} 
[C_i^2(1)+D_i^2(1)]} \right]
\nonumber
\end{eqnarray}
or, equivalently
{\setlength\arraycolsep{0pt}
\begin{eqnarray}
&&\frac{1}{2}{\rm{log}}\Psi_m =\zeta + 
\\
&&{\rm{Re}}\left[\sum_{i=1}^{L}\frac{1}{N_i}\, \frac{\vert\tilde\rho_i\vert\ \gamma_i \;e^{-j\angle \tilde\rho_i}}
{(1+\gamma_i)^2-(\vert\tilde\rho_i\vert \gamma_i)^2}\; \tilde{r}_{i}(k) \, \tilde{r}_{i}^{*}(k-1)e^{-j\frac{2\pi m}{M}}\right]
\nonumber
\end{eqnarray}} 
where $\zeta$ represents the constant term which does not affect the decision.
In (15), the quantities $\vert\tilde\rho_i\vert=\sqrt{\frac{C_i^2(1)+D_i^2(1)}{C_i^2(0)}}$ and $\angle \tilde\rho_i=-\tan^{-1}\left[\frac{D_i(1)}{C_i(1)}\right]$ represent the magnitude and phase of the correlation coefficient $\tilde\rho_i$ given in (8), respectively.

Defining the real-valued weighting factors
\begin{equation}
w_i= \frac{1}{N_i}\, \frac{\vert\tilde\rho_i\vert \gamma_i}{(1+\gamma_i)^2-(\vert\tilde\rho_i\vert \gamma_i)^2},
\end{equation}
it follows from (15) that the optimum combining differential receiver will now compute, for the $k$th symbol, the decision statistics $\left\{\Lambda_{m}(k) \right\}_{m=0}^{M-1}$, and declares that $\Delta\phi(k)=\frac{2\pi n}{M}$ if $\Lambda_{n}(k) = {{\rm{max}}} _m \left\{ \Lambda_{m}(k)\right\}$, where
\begin{equation}
\Lambda_{m}(k)={\rm{Re}}\left[e^{-j\frac{2\pi m}{M}}\,\sum_{i=1}^{L} w_{i}\; \tilde{r}_{i}(k)\,\tilde{r}_{i}^{*}(k-1)\;e^{-j\angle\tilde\rho_i}\right]
\end{equation} 
If the spectrum of the channel complex gain is symmetric, $\tilde\rho_i$ is a real-valued quantity. Then, the optimum combining differential receiver (17) will become [9]
\begin{equation}
\Lambda'_{m}(k)={\rm{Re}}\left[e^{-j\frac{2\pi m}{M}}\,\sum_{i=1}^{L} w_{i}\; \tilde{r}_{i}(k)\,\tilde{r}_{i}^{*}(k-1)\right]
\end{equation} 
If the diversity branches are i.i.d., but the fading gains have asymmetric spectrum, the optimum receiver will become
\begin{equation}
\Lambda''_{m}(k)={\rm{Re}}\left[e^{-j\frac{2\pi m}{M}}\,e^{-j\angle\tilde\rho}\,\sum_{i=1}^{L} \tilde{r}_{i}(k)\,\tilde{r}_{i}^{*}(k-1)\right]
\end{equation} 
where $\tilde\rho=\tilde\rho_i$ for $i=1,2,\cdots,L$. For i.i.d. branches with fading gains having symmetric spectrum, the optimum receiver is the well-known product detector, given by [4]
\begin{equation}
\Lambda'''_{m}(k)={\rm{Re}}\left[e^{-j\frac{2\pi m}{M}}\,\sum_{i=1}^{L}\tilde{r}_{i}(k)\,\tilde{r}_{i}^{*}(k-1)\right]
\end{equation} 
Comparing (20) with (17), we see that in the case of i.n.i.d. channels with asymmetric power spectrum, the receiver first rotates the product phasor $\tilde{r}_{i}(k)\tilde{r}_{i}^{*}(k-1)$ between the two received signal samples at each diversity branch by the angle $-\angle\tilde\rho_i$, then scales each resulting phasor by the weight $w_{i}$, and finally sums all $L$ rotated and scaled phasors to form a decision variable. Clearly, in order to form the optimum detector (17), besides the received signal samples $\tilde{r}_{i}(k)$ and $\tilde{r}_{i}(k-1)$, the receiver requires the {\it a priori} knowledge of the channel statistics, including the power spectral densities of AWGN $N_i$, both the magnitude and phase of the fading correlation coefficient $\tilde\rho_i$, and the mean received SNR $\gamma_i$. These quantities can be pre-computed according to our knowledge of the channel statistics at the receiver.

\section{PERFORMANCE ANALYSIS}

In this section, we will derive exact, explicit and closed-form BEP expressions for differentially detected DPSK for the optimum receiver (17). Due to space limitation, we only consider 8-DPSK in this paper. The signal constellation, bit mapping and the decision region $R_m$ for 8-DPSK is shown in Fig. 2. In [4] and [9], the average BEP is computed using the binary reflected Gray code (BRGC) approach through Hamming weight spectrum [20]. It is shown in [21] that the BRGC approach with Hamming weight is less accurate for $M\geq 16$. In this paper, we adopt a new approach, namely, the average BEP is obtained by calculating the BEP for each of the three individual bits in 8-DPSK. This approach has the advantage of showing explicitly the BEP performance differently for the three different transmitted information bits. Therefore, using the bit which has lower BEP to convey more important information can improve communication reliability. 

From Fig. 2, we see that each signal point is represented by a 3-bit symbol ($j_1,\,j_2,\,j_3$). We use $P_{j_1}$, $P_{j_2}$ and $P_{j_3}$ to denote the corresponding individual BEP. Since the three bits are equally likely, the average BEP is given by
\begin{equation}
P=\frac{1}{3}(P_{j_1}+P_{j_2}+P_{j_3})
\end{equation}
We begin with computing $P_{j_1}$. Without loss of generality, it is assumed that $j_1=0$. The case where $j_1=1$ gives an identical result. From Fig. 2, we see that the bit $j_1=0$ is associated with the symbols 000 $(\Delta\phi(k)=0)$, 001 $(\Delta\phi(k)=\pi/4)$, 011 $(\Delta\phi(k)=\pi/2)$, and 010 $(\Delta\phi(k)=3\pi/4)$. Thus, conditioning on $j_1=0$, the BEP $P_{j_1}$ will be given by
{\setlength\arraycolsep{1pt}
\begin{eqnarray}
P_{j_1}&=&\frac{1}{4}\Big[P_{j_1}(e\vert\Delta\phi(k)=0)+P_{j_1}(e\vert\Delta\phi(k)=\pi/4)
\\
&&+P_{j_1}(e\vert\Delta\phi(k)=\pi/2)+P_{j_1}(e\vert\Delta\phi(k)=3\pi/4)\Big]
\nonumber
\end{eqnarray}} 
Here, $P_{j_1}(e\vert\Delta\phi(k)=m\pi/4),m=0,1,2,3$, is the probability that conditioning on $\Delta\phi(k)=m\pi/4$, the decision $j_1=1$ is made. With reference to Fig. 2, this is equivalent to the probability that conditioning on $\Delta\phi(k)=m\pi/4$, the phasor $\sum_{i=1}^{L} w_{i}\, \tilde{r}_{i}(k)\,\tilde{r}_{i}^{*}(k-1)\,e^{-j\angle\tilde\rho_i}$ lies outside the half-plane region $R_0+R_1+R_2+R_3$ (i.e., in the region $R_4+R_5+R_6+R_7$). The BEP $P_{j_1}(e\vert\Delta\phi(k)=m\pi/4)$ is thus obtained as
{\setlength\arraycolsep{0pt}
\begin{eqnarray}
&&P_{j_1}\big(e\vert\Delta\phi(k)=m\pi/4\big)=P\Bigg\{{\rm Re}\Bigg[ e^{-j\frac{3\pi}{8}}
\\
&&\;\; \times\left(\sum_{i=1}^{L} w_{i}\tilde{r}_{i}(k)
\tilde{r}_{i}^{*}(k-1)e^{-j\angle\tilde\rho_i}\right)\Bigg]<0 
\left. \Bigg\vert\Delta\phi(k)=\frac{m\pi}{4} \right\}
\nonumber
\end{eqnarray}
To evaluate (23), first, it follows from (11) and (12) that conditioning on $\Delta\phi(k)=m\pi/4$ and on $\tilde{r}_{i}(k-1)\,e^{j\angle\tilde\rho_i}\,e^{j3\pi/8} = \tilde\alpha_i,\,{\rm for}\; i=1,2,\cdots,L,$ the quantity $\tilde{r}_{i}(k)$ is conditionally Gaussian with mean $\tilde\alpha_i\,\frac{\tilde\rho_i \gamma_{i}}{1+\gamma_{i}}\, e^{-j\angle\tilde\rho_i}e^{-j3\pi/8}e^{jm\pi/4}=\tilde\alpha_i\,\frac{\vert\tilde\rho_i\vert \gamma_{i}}{1+\gamma_{i}}\, e^{-j3\pi/8}e^{jm\pi/4}$, where $\tilde\rho_i=\vert\tilde\rho_i\vert e^{j\angle\tilde\rho_i}$ has been used, and variance
$\frac{(1+\gamma_{i})^{2}-(\vert\tilde\rho_i\vert\gamma_{i})^{2}}{1+\gamma_{i}}\, N_{i}$. Then, in (23) the quantity ${\rm Re}[ e^{-j\frac{3\pi}{8}}(\sum_{i=1}^{L} w_{i}\tilde{r}_{i}(k) \tilde{r}_{i}^{*}(k-1)e^{-j\angle\tilde\rho_i})]$
is conditionally Gaussian with mean
${\cos{\big(m\pi/4-3\pi/8\big)}}\,\sum_{i=1}^{L} w_i\, \frac{\vert\tilde\rho_i\vert \gamma_{i}}{1+\gamma_{i}}\,\vert \tilde\alpha_i\vert^2$ and variance 
$\frac{1}{2}\sum_{i=1}^{L}w_i^2\,\frac{(1+\gamma_{i})^{2}-(\vert\tilde\rho_i\vert \gamma_{i})^{2}}{1+\gamma_{i}}\, N_{i}\,\vert \tilde\alpha_i\vert^2$. Finally, following the derivation procedure detailed in [9], the BEP in (23) can be obtained as
\begin{eqnarray}
P_{j_1}\left(e\vert\Delta\phi(k)=\frac{m\pi}{4}\right)&=&
\\
&&\hspace{-25pt}\sum_{i=1}^{L}\frac{G_i}{2}
\left[1-\sqrt{\frac{\cos^2\left(\frac{m\pi}{4}-\frac{3\pi}{8}\right)}{\cos^2\left(\frac{m\pi}{4}-\frac{3\pi}{8}\right)+1/\lambda_{i}}}\right]
\nonumber
\end{eqnarray}
where the quantity $G_i$ is given by
\begin{equation}
G_i = \prod_{j=1,j \ne i}^{L}\frac{\lambda_{i}}{\lambda_{i}-\lambda_{j}},\;
{\rm and}\; 
\lambda_{i}= \frac{(\vert\tilde\rho_i\vert \gamma_{i})^2}{(1+\gamma_{i})^{2}-(\vert\tilde\rho_i\vert \gamma_{i})^{2}}
\end{equation}
Putting (24) into (22) leads to the BEP $P_{j1}$. An interesting observation from (24) is that the BEP does not depend on the phase, $\angle\tilde\rho_i$, of the fading correlation coefficient $\tilde\rho_i$. Intuitively, this is because the optimum receiver (17) can provide ``phase compensation'' for each diversity branch before combining using the channel statistic knowledge $e^{-j\angle\tilde\rho_i}$. As such, we expect that the receivers (18) and (20) are suboptimum over the channel with asymmetric fading spectrum.

Next, we compute $P_{j2}$ in (21). The procedure for obtaining the conditional BEP for $j_2=0$ is parallel to that followed in the case for $j_1=0$. From Fig. 2, the bit $j_2=0$ is associated with the symbols 001 $(\Delta\phi(k)=\pi/4)$, 000 $(\Delta\phi(k)=0)$, 100 $(\Delta\phi(k)=7\pi/4)$, and 101 $(\Delta\phi(k)=3\pi/2)$. Hence, conditioning on $j_2=0$, the BEP $P_{j_2}$ is given by
{\setlength\arraycolsep{1pt}
\begin{eqnarray}
P_{j_2}&=&\frac{1}{4}\Big[P_{j_2}(e\vert\Delta\phi(k)=\pi/4)+P_{j_2}(e\vert\Delta\phi(k)=0)
\\
&&+P_{j_2}(e\vert\Delta\phi(k)=7\pi/4)+P_{j_2}(e\vert\Delta\phi(k)=3\pi/2)\Big]
\nonumber
\end{eqnarray}} 
where $P_{j_2}(e\vert\Delta\phi(k)=n\pi/4),n=0,1,6,7$, is the conditional probability that the phasor $\sum_{i=1}^{L} w_{i}\tilde{r}_{i}(k)\tilde{r}_{i}^{*}(k-1)e^{-j\angle\tilde\rho_i}$ lies in the half-plane region $R_2+R_3+R_4+R_5$, i.e.,
{\setlength\arraycolsep{0pt}
\begin{eqnarray}
&&P_{j_2}\big(e\vert\Delta\phi(k)=n\pi/4\big)=P\Bigg\{{\rm Re}\Bigg[ e^{j\frac{\pi}{8}}
\\
&&\;\;\times \left(\sum_{i=1}^{L} w_{i}\tilde{r}_{i}(k)
\tilde{r}_{i}^{*}(k-1)e^{-j\angle\tilde\rho_i}\right)\Bigg]<0 
\left. \Bigg\vert\Delta\phi(k)=\frac{n\pi}{4} \right\}
\nonumber
\end{eqnarray}
which has solution 
\begin{eqnarray}
P_{j_2}\left(e\vert\Delta\phi(k)={n\pi}/{4}\right)&=&
\\
&&\hspace{-30pt}\sum_{i=1}^{L}\frac{G_i}{2}
\left[1-\sqrt{\frac{\cos^2\left(\frac{n\pi}{4}+\frac{\pi}{8}\right)}{\cos^2\left(\frac{n\pi}{4}+\frac{\pi}{8}\right)+1/\lambda_{i}}}\right]
\nonumber
\end{eqnarray}
Putting (28) into (26) leads to the BEP $P_{j2}$.

Finally, we compute $P_{j3}$ in (21). From Fig. 2, the bit $j_3=0$ is associated with the symbols 100 $(\Delta\phi(k)=7\pi/4)$, 000 $(\Delta\phi(k)=0)$,  010 $(\Delta\phi(k)=3\pi/4)$, and 110 $(\Delta\phi(k)=\pi)$. Thus, conditioning on $j_3=0$, the BEP $P_{j_3}$ is given by
{\setlength\arraycolsep{1pt}
\begin{eqnarray}
&&P_{j_3}=\frac{1}{4}\Big[P_{j_3}(e\vert\Delta\phi(k)=7\pi/4)+P_{j_3}(e\vert\Delta\phi(k)=0)
\\
&&\hspace{50pt}+P_{j_3}(e\vert\Delta\phi(k)=3\pi/4)+P_{j_3}(e\vert\Delta\phi(k)=\pi)\Big]
\nonumber
\end{eqnarray}} 
where $P_{j_3}(e\vert\Delta\phi(k)=l\pi/4),l=0,3,4,7$, is the conditional probability that the phasor $\sum_{i=1}^{L} w_{i}\tilde{r}_{i}(k)\tilde{r}_{i}^{*}(k-1)e^{-j\angle\tilde\rho_i}$ lies in the region $R_1+R_2+R_5+R_6$. This is equivalent to the conditional probability that after rotating by $-\pi/8$, the product of the inphase and quadrature-phase components of the phasor $\sum_{i=1}^{L} w_{i}\tilde{r}_{i}(k)\tilde{r}_{i}^{*}(k-1)e^{-j\angle\tilde\rho_i}$ is greater than zero, i.e.,  
{\setlength\arraycolsep{0pt}
\begin{eqnarray}
&&P_{j_3}\big(e\vert\Delta\phi(k)=l\,\pi/4\big)=
\\
&&P\left\{{\rm Re}\left[ e^{-j\frac{\pi}{8}}\left(\sum_{i=1}^{L} w_{i}\tilde{r}_{i}(k)
\tilde{r}_{i}^{*}(k-1)e^{-j\angle\tilde\rho_i}\right)\right] \right.{\rm Im}\Bigg[ e^{-j\frac{\pi}{8}}
\nonumber\\
&&\quad\times\left.\left(\sum_{i=1}^{L} w_{i}\tilde{r}_{i}(k)
\tilde{r}_{i}^{*}(k-1)e^{-j\angle\tilde\rho_i}\right)\right]>0
\left. \Bigg\vert\Delta\phi(k)=\frac{l\,\pi}{4} \right\}
\nonumber
\end{eqnarray}
From the argument for deriving (24), we note that conditioning on $\Delta\phi(k)=l\pi/4$ and on $\tilde{r}_{i}(k-1)\,e^{j\angle\tilde\rho_i}\,e^{j\pi/8} = \tilde\beta_i,\,{\rm for}\; i=1,2,\cdots,L,$
the inphase component in (30), ${\rm Re}[e^{-j\frac{\pi}{8}}(\sum_{i=1}^{L} w_{i}\tilde{r}_{i}(k) \tilde{r}_{i}^{*}(k-1)e^{-j\angle\tilde\rho_i})]$
is conditionally Gaussian with mean
${\cos{\big(l\pi/4-\pi/8\big)}}\,\sum_{i=1}^{L} w_i\, \frac{\vert\tilde\rho_i\vert \gamma_{i}}{1+\gamma_{i}}\,\vert \tilde\beta_i\vert^2$ and variance 
$\frac{1}{2}\sum_{i=1}^{L}w_i^2\,\frac{(1+\gamma_{i})^{2}-(\vert\tilde\rho_i\vert \gamma_{i})^{2}}{1+\gamma_{i}}\, N_{i}\,\vert \tilde\beta_i\vert^2$. Similarly, the component ${\rm Im}[e^{-j\frac{\pi}{8}}(\sum_{i=1}^{L} w_{i}\tilde{r}_{i}(k) \tilde{r}_{i}^{*}(k-1)e^{-j\angle\tilde\rho_i})]$ in (30) is also a conditionally Gaussian random variable, with mean
${\sin{\big(l\pi/4-\pi/8\big)}}\,\sum_{i=1}^{L} w_i\, \frac{\vert\tilde\rho_i\vert \gamma_{i}}{1+\gamma_{i}}\,\vert \tilde\beta_i\vert^2$ and variance 
$\frac{1}{2}\sum_{i=1}^{L}w_i^2\,\frac{(1+\gamma_{i})^{2}-(\vert\tilde\rho_i\vert \gamma_{i})^{2}}{1+\gamma_{i}}\, N_{i}\,\vert \tilde\beta_i\vert^2$. Moreover, it follows from (13) and the properties of the complex Gaussian random variables [18] that the conditional inphase and quadrature-phase components ${\rm Re}[e^{-j\frac{\pi}{8}}(\sum_{i=1}^{L} w_{i}\tilde{r}_{i}(k) \tilde{r}_{i}^{*}(k-1)e^{-j\angle\tilde\rho_i})]$ and ${\rm Im}[e^{-j\frac{\pi}{8}}(\sum_{i=1}^{L} w_{i}\tilde{r}_{i}(k) \tilde{r}_{i}^{*}(k-1)e^{-j\angle\tilde\rho_i})]$ in (30) are also independent. Therefore, conditioning on $\Delta\phi(k)=l\pi/4$ and on $\tilde{r}_{i}(k-1)\,e^{j\angle\tilde\rho_i}\,e^{j\pi/8} = \tilde\beta_i$, and denoting the inphase and quadrature-phase components as
\begin{eqnarray}
X \;&\sim&\; N\Big({\cos{\big(l\pi/4-\pi/8\big)}}\,u, \;\eta^2 \Big)
\nonumber\\
Y \;&\sim&\; N\Big({\sin{\big(l\pi/4-\pi/8\big)}}\,u, \;\eta^2 \Big)
\end{eqnarray}
where $u$ and $\eta^2$ are given, respectively, by
\begin{eqnarray}
u &=& \sum_{i=1}^{L} w_i\, \frac{\vert\tilde\rho_i\vert \gamma_{i}}{1+\gamma_{i}}\,\vert \tilde\beta_i\vert^2
\nonumber\\
\eta^2 &=& \frac{1}{2}\sum_{i=1}^{L}w_i^2\,\frac{(1+\gamma_{i})^{2}-(\vert\tilde\rho_i\vert \gamma_{i})^{2}}{1+\gamma_{i}}\, N_{i}\,\vert \tilde\beta_i\vert^2
\end{eqnarray}
the conditional BEP $P_{j_3}\big(e\big\vert\Delta\phi(k)=\frac{l\pi}{4}, \tilde\beta_i\big)$ is given by
\begin{equation}
P_{j_3}\big(e\big\vert\Delta\phi(k)=\frac{l\pi}{4}, \tilde\beta_i\big)
=P\left(X\,Y>0\big\vert\Delta\phi(k)=\frac{l\pi}{4}, \tilde\beta_i\right).
\end{equation}
This is probability that the product of two independent real-valued Gaussian random variables with non-zero, nonidentical means and identical variances is greater than zero. This is a special case of the results given in [2, Appendix B] concerning the probability that a general quadratic form in complex-valued Gaussian random variables is less than zero. Using [2, (B-21) of Appendix B], (33) can be evaluated as
{\setlength\arraycolsep{0pt}
\begin{eqnarray}
&&P_{j_3}\big(e\big\vert\Delta\phi(k)=\frac{l\pi}{4},\tilde\beta_i\big)=1-
\\
&&Q_1\Big(\sqrt{g[1-\sin{(l\pi/2-\pi/4)}]},\sqrt{g[1+\sin{(l\pi/2-\pi/4)}]}\Big)
\nonumber\\
&&+\frac{1}{2}I_0\left[g\vert\cos{(l\pi/2-\pi/4)}\vert\right]\exp(-g)
\nonumber
\end{eqnarray}} 
where, $Q_1(a,b)$ is first-order Marcum's $Q$-function and $I_k(x)$ is the $k$th-order modified Bessel function of the first kind. In (34), the quantity $g=\sum_{i=1}^{L}w'_{i}\,\vert \tilde\beta_i\vert^2$ has PDF given by [9]
\begin{equation}
p(g)=\sum_{i=1}^{L}\frac{G_i}{w'_{i}\, N_i\, (1+\gamma_{i})}{\rm{exp}}\left[-\frac{g}{w'_{i}\, N_i\, (1+\gamma_{i})}\right]
\end{equation}
where
$
w'_{i}=\frac{1}{N_i}\,\frac{(\vert\tilde\rho_i\vert \gamma_{i})^2}{(1+\gamma_{i})\left[(1+\gamma_{i})^{2}-(\vert\tilde\rho_i\vert \gamma_{i})^{2}\right]}.
$
Averaging the conditional probability (34) over $g$ using the PDF (35) gives the BEP 
$P_{j_3}\big(e\big\vert\Delta\phi(k)={l\pi}/{4}\big)$ in (30), i.e.,
\begin{equation}
P_{j_3}\big(e\big\vert\Delta\phi(k)=\frac{l\pi}{4}\big)=\int_0^\infty
P_{j_3}\big(e\big\vert\Delta\phi(k)=\frac{l\pi}{4},\tilde\beta_i\big)\,p(g) dg
\end{equation}
Substituting (34) and (35) into (36), we obtain, after manipulation and simplification, 
{\setlength\arraycolsep{-3pt}
\begin{eqnarray}
&&P_{j_3}\big(e\big\vert\Delta\phi(k)=\frac{l\pi}{4}\big)=\sum_{i=1}^{L}\frac{G_i}{\lambda_i}\Bigg[
\frac{1}{\sqrt{A_i^2-\cos^2{(\frac{l\pi}{2}-\frac{\pi}{4})}}}
\\
&&\times \frac{1}{1-\frac{\vert\cos{(\frac{l\pi}{2}-\frac{\pi}{4})}\vert(\sqrt{2}-1)}{A_i+
\sqrt{A_i^2-\cos^2{(\frac{l\pi}{2}-\frac{\pi}{4})}}}}-\frac{1/2}{{\sqrt{A_i^2-
\cos^2{(\frac{l\pi}{2}-\frac{\pi}{4})}}}}\Bigg]
\nonumber
\end{eqnarray}
where $A_i$ is given by
\begin{eqnarray}
A_i=\left(\frac{1+\gamma_{i}}{\vert\tilde\rho_i\vert \gamma_{i}}\right)^2
\nonumber
\end{eqnarray}
Putting (37) into (29) leads to the BEP $P_{j3}$. Substituting (22), (26) and (29) in (21), we obtain the average BEP $P$.

\section{NUMERICAL EXAMPLE}

Fig. 3 plots the BEP performance for the three individual bits in (22), (26) and (29) and the average BEP in (21) of 8-DPSK against the total average received SNR per bit. The order of diversity is set to $L=2$. The abscissa represents the total mean SNR per bit which is given by $\gamma_b=\sum_{i=1}^{2}\gamma_{i}^b=\frac{1}{3}\sum_{i=1}^{2}\gamma_{i}$. The average received bit energy distribution among the two branches is set to $\gamma_1^b:\gamma_2^b=30\%:70\%$. It is assumed that the fading correlation coefficient (the normalized covariance function) model follows [14, eq.(10)], given by
\begin{equation}
\frac{{\rm{E}}[\tilde{c}_{i}(t)\tilde{c}_{i}^{*}(t-\tau)]}{{\rm{E}}\big[\vert\tilde{c}_{i}(t)\vert^2\big]}=
\frac{I_0\left(\sqrt{\kappa^2-4\pi^2f_d^2\tau^2+j4\pi\kappa f_d\tau}\right)}{I_0(\kappa)}
\end{equation}
where $f_d$ is the Doppler frequency, and $\kappa$ is a parameter that controls the width of the angle of arrival of scatter components [14, eq.(1)]. Note that if $\kappa=0$, (38) results in the correlation coefficient for the Jakes two-dimensional isotropic scattering model, i.e., ${{\rm{E}}[\tilde{c}_{i}(t)\tilde{c}_{i}^{*}(t-\tau)]}/{{\rm{E}}\big[\vert\tilde{c}_{i}(t)\vert^2\big]}=I_0(j2\pi f_m\tau)=J_0(2\pi f_m\tau)$, where $J_0(\cdot)$ is the zeroth-order Bessel function. 
We assume that the normalized Doppler spread $f_dT=0.03$ and $0.05$ for diversity branches 1 and 2, respectively, and the parameter $\kappa$ is set to $3$. Thus, we have $\tilde\rho_1=0.9871+j0.1519$ and $\tilde\rho_2=0.9642+j0.2511$. It is seen from Fig. 3 that the third bit $j_3$ has the lowest BEP, whereas, the BEP $P_{j_1}$ for the first bit $j_1$ is equal to the BEP $P_{j_2}$ for the second bit $j_2$.

\begin{figure}[h]
\center
\includegraphics[height=0.1\textwidth,width=0.45\textwidth]
{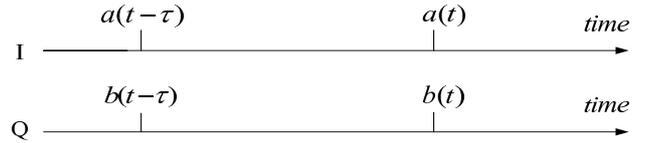}
\caption{Illustration of complex channel fading process.}
\end{figure}
\begin{figure}[ht]
\center
\includegraphics[height=0.4\textwidth,width=0.45\textwidth]
{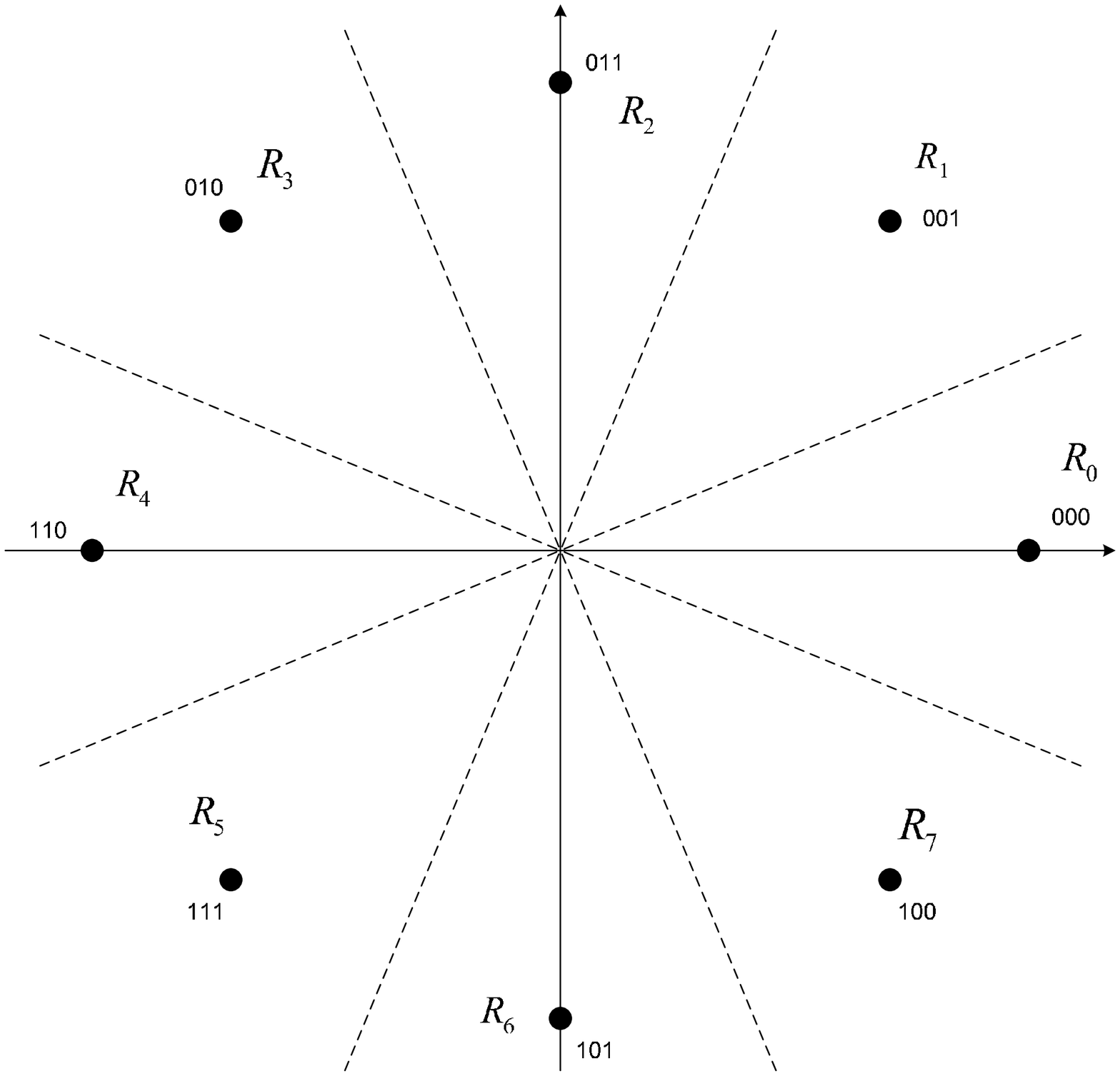}
\caption{8-DPSK constellation and decision region.}
\end{figure}
\begin{figure}[h]
\centerline{\epsfxsize=8.5cm\epsfysize=7.0cm\epsffile{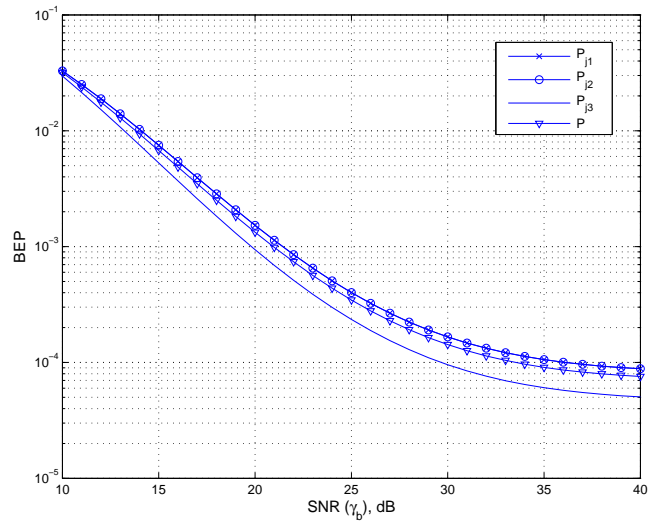}}
\caption{BEP comparison of the three individual bits and the average of all bits for 8-DPSK.}
\end{figure}

\begin{thebibliography}{11} 
\bibitem{ref:1}
M. Schwartz, W.R. Bennett, and S. Stein,
{\em Communication Systems and Techniques},
New York: McGraw-Hill, 1966.
\bibitem{ref:2}
J.~G. Proakis,
{\em Digital Communications},
4{th} edition, New York: McGraw-Hill, 2001.
\bibitem{ref:3}
M.~K. Simon and M.~S. Alouini,,
{\em Digital communication over fading channels}, 2nd Edition,
New York: John Wiley \& Sons, 2005.
\bibitem{ref:4}
P.~Y. Kam,
``Bit error probabilities of MDPSK over the nonselective Rayleigh fading channel with diversity reception,"
{\em IEEE Trans. Commun.},
vol.39, pp.220-224, February 1991.
\bibitem{ref:5}
M.Z. Win and J.H. Winters,
``Analysis of hybrid selection/maximal-ratio combining of diversity channels with unequal SNR in Rayleigh fading,''
{\em Proc. 49th IEEE VTC},
pp.215-220, May 16-20, 1999.
\bibitem{ref:6}
P. Polydorou and P. Ho,
``Error performance of MPSK with diversity combining in non-uniform Rayleigh fading and non-ideal
channel estimation,''
{\em Proc. 51st IEEE VTC},
pp.627-631, May 15-18, 2000.
\bibitem{ref:7}
F. Adachi,
``Postdetection optimal diversity combiner for DPSK differential detection,''
{\em IEEE Trans. Veh. Technology},
vol.42, pp.326-337, August 1993.
\bibitem{ref:8}
H. Fu and P.~Y. Kam,
``Performance of Optimum and Suboptimum Combining Diversity
Reception for Binary DPSK over Independent, Nonidentical Rayleigh
Fading Channels,"
{\em Proc. 40th IEEE ICC},
pp.2367-2371, May 16-20, 2005.
\bibitem{ref:9}
H. Fu and P.~Y. Kam,
``MDPSK diversity receiver over Rayleigh fading channels with
differential detection and nonidentical branch statistics,"
{\em Proc. 63rd IEEE VTC},
pp.1660-1664, May 7-10, 2006.
\bibitem{ref:10}
H. Fu and P.~Y. Kam,
``Performance of Optimum and Suboptimum Combining Diversity
Reception for Binary and Quadrature DPSK over Independent, Nonidentical Rayleigh
Fading Channels,"
to appear in the {\em IEEE Trans. Commun.}, May 2007.
\bibitem{ref:11}
L.~J. Mason
``Error probability evaluation for systems employing differential detection in a
Rician fast fading environment and Gaussian noise,''
{\em IEEE Trans. Commun.},
vol.35, pp.39-46, January 1987.
\bibitem{ref:12}
W.C. Jakes, 
{\em Microwave Mobile Communications},
NJ: IEEE Press, 1974.
\bibitem{ref:13}
M. Patzold, Y.C. Li and F. Laue 
``A study of a land mobile satellite channel model with asymmetrical Doppler power spectrum and
lognormally distributed line-of-sight component,''
{\em IEEE Trans. Veh. Technology},
vol.47, pp.297-310, February 1998.
\bibitem{ref:14}
A. Abdi, J.A. Barger and M. Kaveh 
``A parametric model for the distribution of the angle of arrival and the associated correlation function and power spectrum at the mobile station,''
{\em IEEE Trans. Veh. Technology},
vol.51, pp.425-434, May 2002.
\bibitem{ref:15}
K. Anim-Appiah
``Complex envelope correlations for nonisotropic scattering,''
{\em Electron. Lett.},
vol.34, pp.918-919, April 1998.
\bibitem{ref:16}
W.~B. Davenport and W.~L. Root,
{\em An introduction to the theory of random signals and noise},
New York: McGraw-Hill, 1958 
\bibitem{ref:17}
F.~D. Neeser and J.L. Massey
``Proper complex random processes with applications to information theory,''
{\em IEEE Trans. Inform. Theory},
vol.39, pp.1293-1302, July 1993.
\bibitem{ref:18}
S.M. Kay,
{\em Fundamentals of Statistical Signal Processing: Estimation Theory},
New Jersey: Prentice-Hall, 1998.
\bibitem{ref:19}
A. Papoulis and S.U. Pillai,
{\em Probability, Random Variables and Stochastic Processes},
4th Ed., MA: McGraw-Hill, 2002.
\bibitem{ref:20}
P.~J. Lee,
``Computation of the bit error rate of coherent M-ary PSK with Gray code bit mapping,''
{\em IEEE Trans. Commun.},
vol.34, pp.488-491, May 1986.
\bibitem{ref:21}
J. Lassing, E.G. Strom, E. Agrell and L. Ottosson,
``Computation of the exact bit-error rate of coherent M-ary PSK with Gray code bit mapping,''
{\em IEEE Trans. Commun.},
vol.51, pp.1758-1760, November 2003.
\end{thebibliography}
\end{document}